\documentclass[aps,prl,showpacs,preprintnumbers,amsmath,amssymb,floatfix,reprint]{revtex4-1}
\usepackage{graphicx}
\usepackage{bm}
\usepackage{times}
\usepackage{color}

\newcommand{\ie}{{\em i.e., }}
\newcommand{\mr}[1]{\mathrm{#1}}
\newcommand{\fpath}{f_{\text{in}}}
\newcommand{\fpathk}{f_{\text{in}}^{(k)}}

\newcommand{\Ft}{{\mathbf{F}_{\!t}}}

\begin{document}
\title{Phase-Field Reaction-Pathway Kinetics of Martensitic Transformations in a Model Fe${}_3$Ni Alloy}
\author{Christophe Denoual}
\email{christophe.denoual@cea.fr}
\author{Anna Maria Caucci}
\author{Laurent Soulard}
\author{Yves-Patrick Pellegrini}

\affiliation{CEA, DAM, DIF, F--91297 Arpajon, France.}

\date{15 July 2010}

\begin{abstract}
A three-dimensional phase-field approach to martensitic transformations that uses reaction pathways in place of a Landau potential is introduced and applied to a model of Fe${}_3$Ni. Pathway branching involves an unbounded set of variants through duplication and rotations by the rotation point groups of the austenite and martensite phases. Path properties, including potential energy and elastic tensors, are calibrated by molecular statics. Acoustic waves are dealt with via a splitting technique between elastic and dissipative behaviors in a large-deformation framework. The sole free parameter of the model is the damping coefficient associated to transformations, tuned by comparisons with molecular dynamics simulations. Good quantitative agreement is then obtained between both methods.
\end{abstract}
\pacs{64.60.--i, 64.70.K--, 81.30.Kf}
\maketitle
Nanoscale materials that undergo martensitic transformations (MT) bear the promise of an exceptional technological revolution \cite{BHAT05}. MTs are displacive structural transitions associated to large inelastic strains, that occur under temperature or loading changes, from a high- (austenite) to a low-symmetry state (martensite) declined in a number of ``variants" \cite{NISH78}, of time scale down to subnanosecond order \cite{KADA03,YAAK05}. Bulk MTs in large samples lead to complex microstructures, due to competing long-range elasticity, and crystallographic constraints on variants \cite{NISH78}. Kinetics of MTs has been investigated at small scales by molecular dynamics (MD) (e.g., \cite{KADA02}), whereas continuum-mechanics-based phase-field (PF) models \cite{WANG97,LEVI09} must be used for large sizes and simulation durations.

The unsolved issue addressed in this Letter consists in seeking quantitative agreement between PF, and MD in its operative range of size- and time-scales, in a time-dependent setting. We focus on the illustrative case of strain-driven transformations near 0 K in a stoichiometric (ordered) Fe${}_3$Ni alloy stable at low temperatures only \cite{MOFF84,MEYE98}, that undergoes a proper (i.e., with no shuffling) austenite $\gamma$(fcc) $\to$ martensite $\alpha$(bcc) transformation along a path of homogeneous deformation of the unit cell. Consistently benchmarking PF calculations by MD simulations requires adjusting the PF model using the empirical potential of the simulations, instead of more accurate first-principles methods (e.g., \cite{MISH05}). Disregarding magnetic degrees of freedom, we use a Meyer-Entel EAM potential developed to investigate the phase diagram of the MT transition in Fe${}_x$Ni${}_{1-x}$ alloys \cite{MEYE98,KADA03} (but see also \cite{MISH05}).

\begin{figure}[h!]
\includegraphics[width=8.5cm]{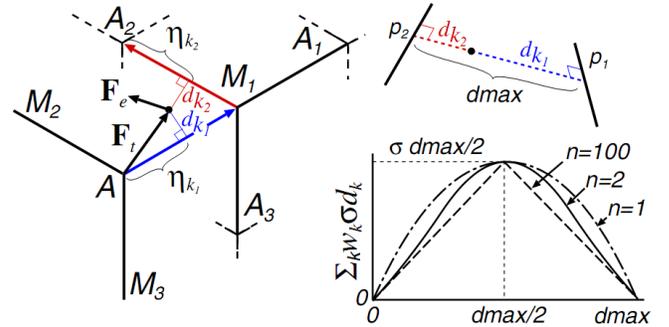}
\caption{\label{fig:fig1} (color online). Left: cycling transformations austenite A $\to$ martensite M $\to$ A, etc., produces in strain gradient ($\mathbf{F}$) space an infinite backbone of preferred RPs for the transformational strain $\Ft$. However, $\Ft$ can depart from it during out-of-path transitions. Right: Inelastic energy along the line (dash) that minimizes the distance between two nonconnected pathways, in a hypothetic pathway arrangement chosen for ease of representation (top). The role of $\sigma$ and $\eta$ are emphasized, inelastic energies on pathways $\widetilde{f}_{\rm in}^{(k)}(\eta_k)$, constant in this case, being set to 0. The parameter $\sigma$ scales the barrier energy, taken proportional to the distance between RPs, and $n$ is an empirical shape parameter (bottom), see Eqs.~(\ref{e:outsideRP}) and (\ref{e:RP_switch}). Thus, direct transitions between remote RPs are inhibited.}
\end{figure}

In the PF method for proper MTs, the nonrelaxed Helm\-holtz energy density has been modeled by a Landau potential for the total strain. Levitas \textit{et al.}\ recently extended this formulation to large strains, using a vector order parameter $\bm{\eta}$ associated to a Landau potential that describes the transformational part of the strain \cite{LEVI09}. However, due to group-subgroup relations in the lattice symmetry point group (PG), the fcc $\to$ bcc transformation is \emph{reconstructive} \cite{BHAT04}. That is, once a martensite variant is reached from the parent phase, different austenite variants, among which the original one, can be reached in turn from the martensite (Fig.\ \ref{fig:fig1}). Repeatedly applying PG transformations thus implies considering an infinite set of variants. Landau theory is then inapplicable, though an approximate theory with non-commensurable order parameter can be used \cite{BHAT04}.

In this context, as a simple alternative to using Landau potentials, we introduce a PF approach based on \emph{reaction pathways} (PF-RP). The RP is a minimum-energy path that links two (meta)stable states via a saddle point (e.g., \cite{SAND09}). Consider two states, austenite (A) and martensite (M), of deformation gradients $\mathbf{F}\equiv\mathbf{I}+\bm{\nabla}\mathbf{u}=\mathbf{F}^{\text{A},\text{M}}$ ($\mathbf{I}$ is the identity and $\mathbf{u}$ is the material displacement) with respect to an arbitrary reference state, which are local energy minimizers with associated elastic moduli tensors $\mathbb{C}^{\text{A},\text{M}}$, computed by molecular statics (MS). 
Although the actual path might slightly differ \cite{SAND09}, we approximate the RP between these states by optimizing over volume \cite{CASP04} along a Bain path \cite{NISH78} $\mathbf{F}^{\text{B}}(s)=s\mathbf{F}^{\text{A}}+(1-s)\mathbf{F}^{\text{M}}$, where $0\leq s\leq 1$ is a path coordinate. Thus along the RP, the Helmholtz energy density is $\widetilde{f}(s)=\min_{k>0}f^{\text{MS}}(k\mathbf{F}^{\text{B}}(s))$ where $f^{\text{MS}}$ is the energy density from MS, associated to the deformation gradient $\widetilde{\mathbf{F}}(s)=k(s)\,\mathbf{F}^{\text{B}}(s)$ where $k(s)$ is the volume-minimizing factor. To comply with crystal symmetry, this first RP is duplicated and rotated for any rotations $\mathbf{R}$ of the austenite and martensite point groups (PG). Because of the high symmetry of the considered phases, only 3 of the 24 possible rotations lead to new RPs (e.g., rotations $\mathbf{R}^{\left<100\right>}_{90^\mr{o}}$, $\mathbf{R}^{\left<010\right>}_{90^\mr{o}}$, $\mathbf{R}^{\left<001\right>}_{90^\mr{o}}$ for austenite, $\mathbf{R}^{\left<110\right>}_{90^\mr{o}}$, $\mathbf{R}^{\left<1\bar{1}0\right>}_{90^\mr{o}}$, $\mathbf{R}^{\left<001\right>}_{90^\mr{o}}$ for one of the first martensites, \cite{NISH78}). As a substitute for the Landau potential of the total strain, we gather these RPs in a backbone construct for the potential. Limiting ourselves to moderate deformations (up to $50\%$), we consider only the first 21 variants \cite{CASP04}. The reference state is the austenite denoted by $\text{fcc}$ where $\mathbf{F}^A=\mathbf{I}$. This backbone captures the most important information about energy barriers.

Quite generally, $\widetilde{f}(s)$ embodies the elastic energy near its minima $s=0,1$, so that $s$ is not exclusively related to transformational behavior.
To deal with a large-deformation theory where Landau-like parameters are associated to a transformational strain \cite{LEVI09}, we adapt a small-strain procedure (introduced in \cite{DENO04} and discussed in \cite{PELL09}) allowing one to derive a transformation-related potential that excludes linear-elastic energy. An internal transformation variable $\eta$ (the phase-field) is introduced by defining the transformational gradient along the RP to be $\widetilde{\mathbf{F}}_t(\eta)\equiv \widetilde{\mathbf{F}}(s(\eta))$, where the function $\eta(s)$ that provides by inversion the relationship $s=s(\eta)$ in this definition must be determined. Adopting the usual large-deformation multiplicative composition law, define the elastic gradient as $\mathbf{F}_e(s,\eta)=\widetilde{\mathbf{F}}(s)\cdot\widetilde{\mathbf{F}}_t^{-1}(\eta)$, and write the energy density along the RP in the alternative form
\begin{equation}
\label{e:fhelRP}
f(s,\eta)=\textstyle{\frac{1}{2}}\mathbf{E}_e(s,\eta):\mathbb{C}(s):\mathbf{E}_e(s,\eta)+\fpath(\eta).
\end{equation}
The first term is an elastic energy expressed using the Green-Lagrange elastic strain $\mathbf{E}_e=\frac{1}{2}\left(\mathbf{F}_e^T\cdot\mathbf{F}_e-\mathbf{I}\right)$, and $\fpath$ is the inelastic energy of the RP \cite{DENO04}. To avoid brutal elastic variations, the elastic tensor $\mathbb{C}$ is made $s$-dependant along the path and taken as a cubic interpolation between $\mathbb{C}^{\text{A,M}}$ with  $\mathbb{C}'(s) =0$  at both ends \cite{LEVI09}. Functions $\fpath(\eta)$ and $\eta(s)$ are then obtained by imposing an exact equality between $\widetilde{f}(s)$ and $f(s,\eta)$ in a relaxed state where $\eta$ is adiabatically eliminated \cite{PELL09}, which yields the necessary equations to be solved numerically:
\begin{equation}
\widetilde{f}(s)=f(s,\eta),\qquad\partial_{\eta}f(s,\eta)=0.
\end{equation}
%
%
Similarly, a mechanism whereby a phase strain can leave a RP to rejoin a neighboring one has to be added \cite{LEVI09}. A transformation gradient $\mathbf{F}_{\!t}$ outside RPs is introduced as a generalized internal variable and is associated to a new potential. Keeping in mind that this potential corresponds to transient states between two arbitrary RPs, (i.e., $f_{\rm in}$ should be a function of $\mathbf{F}_{\!t}$) the energy to be added to $\widetilde{f}_\mr{in}^{(k)}(\eta_k)$ is simply chosen as proportional to the distance $d_k(\Ft) = \min_\eta|\Ft \cdot \widetilde{\mathbf{F}}^{(k)\,-1}_{\!t}(\eta_k)-\mathbf{I}|$ from the $k^{\text{th}}$ RP, where $|\mathbf{A}|=(A_{ij}A_{ij})^{1/2}$ (Fig.~\ref{fig:fig1}, left). The contribution for one RP introduces one parameter $\sigma$ (to be fitted by MS)
\begin{equation}
\label{e:outsideRP}
\fpathk(\mathbf{F}_{\!t})=\widetilde{f}_\mr{in}^{(k)}(\eta_k)+ \sigma d_k(\mathbf{F}_{\!t}) \; ,
\end{equation}
where $\eta_k$ is the argmin in $d_k(\mathbf{F}_{\!t})$, a rotation-independent (i.e., objective) function.

For the complete pathway tree, the overall inelastic energy is an interpolation between potentials
\begin{equation}
\label{e:RP_switch}
f_\mr{in}(\mathbf{F}_{\!t})=\sum_k w_k(\mathbf{F}_{\!t}) f_\mr{in}^{(k)}(\mathbf{F}_{\!t})\;,
\end{equation}
with a partition of unity $w_k$ chosen so that a RP dominates its immediate surrounding, \ie $w_k =1$ for $\Ft$ near the $(k)$th RP. A simple and convenient choice is to use a function of the distance $d_k$ defined above: $w_k = d^{-n}_k / \sum_i d^{-n}_i$ with  $n>0$ controling the transition between pathways. Whereas $n\rightarrow\infty$ makes $f_\mr{in}$ switch to the nearest RP, best agreement with MS is obtained using $n\approx 2$, which provides smoother transitions (Fig.~\ref{fig:fig1}, right; see also \cite{EPAPS}). The full potential now reads
\begin{equation}
\label{e:fhelRP2}
f(\mathbf{F},\mathbf{F}_{\!t})=\textstyle{\frac{1}{2}}\mathbf{E}_e(\mathbf{F},\mathbf{F}_{\!t}):\mathbb{C}(\mathbf{F}):\mathbf{E}_e(\mathbf{F},\mathbf{F}_{\!t})+\fpath(\mathbf{F}_{\!t}),
\end{equation}
where $\mathbb{C}(\mathbf{F})$ is interpolated from the $\mathbb{C}(s_k)$s, using an equation similar to Eq.~(\ref{e:RP_switch}), with $d_k(\mathbf{F}) = \min_{s_k}|\mathbf{F} \cdot \widetilde{\mathbf{F}}^{(k)\,-1}(s_k)-\mathbf{I}|$ to define $s_k$. Along RP $k$, we have $w_k=1$ and $w_{i\neq k}=0$, and Eq.~(\ref{e:fhelRP2}) is equivalent to Eq.~(\ref{e:fhelRP}). For transition between pathways, the energy barrier is proportional to the distance between RPs, which naturally inhibits unphysical transitions between ``distant'' variants (Fig.~\ref{fig:fig1}, right). This approach is quite different from the interpolation scheme used in \cite{LEVI09} and, we believe, simpler to handle, at least for reconstructive transformations involving an extended reaction tree.

By construction, linear-elastic energy is removed from $\fpath(\Ft)$ which, apart from small nonlinear elastic contributions, describes the non-convex (unstable) part of the energy along the RP \cite{PELL09}. Hence, damping can be prescribed for $\Ft$ while leaving linear-elastic wave dynamics undamped, consistently with the negligible character of viscoelasticity in solid metals. With $f$ given by (\ref{e:fhelRP2}), $\Ft$ follows by hypothesis a time-dependent Ginzburg-Landau kinetics of parameter $\nu$:
\begin{equation}
\label{e:TDGL}
\dot{\mathbf{F}}_{\!t}=-\nu^{-1}\partial_{\mathbf{F}_{\!t}} f(\mathbf{F}, \Ft).
\end{equation}
In $\nu$ are lumped dissipation mechanisms such as vibrational or magnetism entropy \cite{Delaire_etal_05}, nonlinear acoustic waves, as well as couplings to inessential lattice degrees of freedom that were adiabatically eliminated when computing $\widetilde{f}(s)$.

Finally, the dynamics of $\mathbf{u}$ obeys the equation $\rho\ddot{\mathbf{u}}=\bm{\nabla}\cdot\bm{\sigma}$, where $\mathbf{\rho}$ is the local density, and where the Cauchy stress $\bm{\sigma}$ is related to the first Piola-Kirchhoff stress $\mathbf{P}=\partial f(\mathbf{F},\mathbf{F}_{\!t})/\partial \mathbf{F}$. The model is implemented in a Lagrangian code using an element-free Galerkin (EFG) formulation \cite{BELY94} in total strain, with explicit time integration in a form able to handle acoustic wave propagation and rapid phase changes \cite{ZIEN00}. This least-square formulation of EFG produces smooth fields for $\mathbf{u}$ and $\Ft$ with no pinning at interpolation nodes. Hence, including an interface-penalizing (gradient) term in the energy is not necessary, the same overall effect being obtained by keeping finite the distance between interpolation nodes. Contrary to $\sigma$ and $n$, the free parameter $\nu$ must be fitted on global simulations.

A benchmark MD simulation is conducted with initial temperature $T=0$ K. An austenitic cube of size $L=30 a_0$ with lattice parameter $a_0=3.64$ {\AA} ($108\,000$ atoms) is deformed with time $t\leq t_{\text{f}}$, with $t_{\text{f}}=80$ ps, according to a time-dependent overall strain $\overline{\mathbf{F}}(t) = \mathbf{I}+(t/t_{\text{f}})[\alpha \mathbf{F}^{\text{M}_1}+(1-\alpha)\mathbf{F}^{\text{M}_2}-\mathbf{I}]$ with components of $\mathbf{F}^{\text{M}_1} = (1.105;1.105;0.789)$ and $\mathbf{F}^{\text{M}_2} = (0.789;1.105;1.105)$ and $\alpha$ ranging from 0.1 to 0.3 to obtain various compounds of two martensite variants $\text{M}_{1,2}$ in the final configuration. Periodic boundary conditions (PBCs) are used. At $t=t_{\text{f}}$, acoustic waves have run 26 times across the sample, which evenly spreads out perturbations.

For the corresponding PF-RP simulations, node spacing controls the spatial resolution. A good compromise between resolution and computational cost is obtained with  27 atoms per interpolation node (4000 nodes) placed in the same fcc arrangement as that used for MD. Values $\sigma=1.9$~GPa and $n=2.2$ lead to a good reproduction of the energy between two martensite variants $M_1$ and $M_2$, for prescribed strains $\mathbf{F}=\alpha\mathbf{F}^\mr{M_1}+(1-\alpha)\mathbf{F}^\mr{M_2}$ with $\alpha$ ranging between 0 and 0.5 (see \cite{EPAPS}).

\begin{figure}
\includegraphics[width=8.5cm]{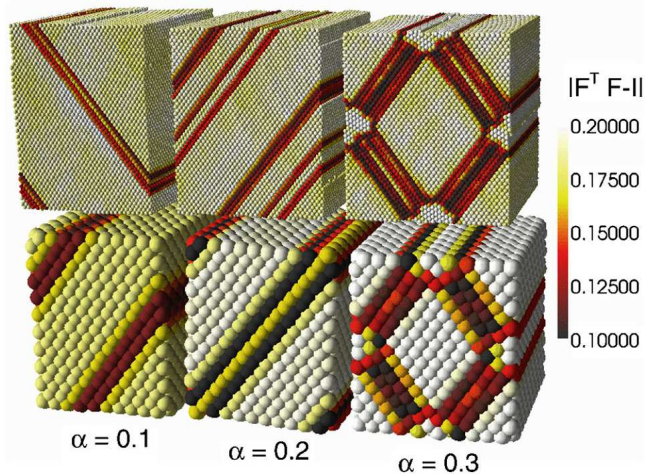}
\caption{\label{fig:fig2} (color online). Comparison of deformation measure between MD (top) and PF-RP (bottom) for $\alpha=\{0.1, 0.2, 0.3\}$ at 80 ps (light: martensite; dark: incompletely-transformed austenite).}
\end{figure}

\begin{figure}
\includegraphics[width=8.5cm]{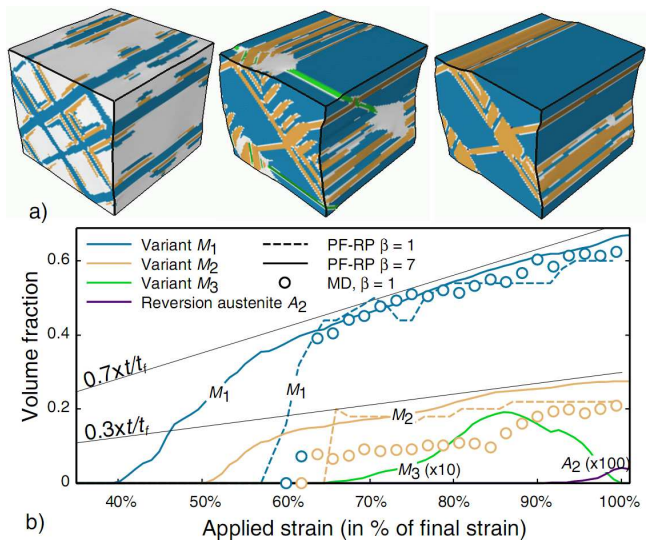}
\caption{\label{fig:fig3} (color online). (a)
Intermediate and final microstructures for $\beta=7$. (b) Volume fraction of variants for small ($\beta = 1$) and large ($\beta = 7$) simulations for MD (dots) and PF-RP (plain and dashed lines). Simple composition rule between initial austenite and variants is plotted (black lines). For $\beta=7$, variant 3 appears as an intermediate phase. A small fraction of ``reversion'' austenite (label A2) is produced at variant angle points.}
\end{figure}

For PF-RP and MD, $\mathbf{F}$ is used to monitor deformation, and to identify variants. For MD it is obtained from displacements $\mathbf{u}$ of neighboring atoms, by least-square optimization. Final states in MD are made of bands of martensite for $\alpha \leq 0.2$ whereas ``chessboard''-like structures \cite{BHAT99} emerge for $\alpha=0.3$. For PF-RP, the viscosity $\nu$ controls the final states, a very good match between MD and PF-RP being obtained with $\nu \leq14$ mPa.s (Fig.~\ref{fig:fig2}). Lowering $\nu$ does not change the final states, but increasing it turns the bandlike structures into chessboard ($0.014<\nu < 0.06$), then into a homogeneously deformed state ($\nu \geq 0.06$). Phase volume fraction are monitored, see Fig.\ref{fig:fig3}(b). The time to nucleation (TTN) is delayed by increasing $\nu$, best match between MD and PF-RP TTNs being obtained with again $\nu=14$ mPa.s.

This value of $\nu$ is used in all calculations below. It is within a factor $\sim2.5$ of Fe and Ni viscosities at their melting point, close to one another (5.8~mPa.s and 5.4~mPa.s, respectively \cite{BEYE72}). An explanation may be that configurational changes associated to barrier-crossing involve large atom motions comparable to that encountered on the liquidus (even though martensitic transformations are non-diffusive), with similar damping effects.

Analogous chessboard patterns have been observed and reproduced by PF in connection with diffusive alloy decomposition \cite{LEBO98}. For displacive transformations, these ubiquitous structures \cite{BHAT99} have also been obtained in two-dimensional PF calculations: very unstable, they are stabilized in small samples, but decay into more conventional laminatelike structures at large sizes \cite{BOUV09}. We make here a first exploration of this physically important effect in three dimensions (3D) using PF-RP for an imposed deformation $\overline{\mathbf{F}}(t)$ with $\alpha=0.3$. Because of inertial dynamics, convergence of strains to stable values is limited by wave propagation. Therefore, the ratio $t_{\text{f}}/L$ between simulation time $t_{\text{f}}$ and sample size $L$ must be kept constant to allow for meaningful comparisons. Taking the size and duration of the previous calculation as a reference ($\beta=1$), $L$  and $t_{\text{f}}$ are increased by factors $\beta=2$ to $\beta=7$. As deformation proceeds and for $\beta\geq 3$, the initial austenite goes through a chessboard state that decays, via an intermediate mixed structure, to a complex three-dimensional laminate state of austenite mixed with twin bands of martensite. This sequence is illustrated on Figs.\ \ref{fig:fig3}(a) for the largest size $L= 210\, a_0$ ($\beta=7$) and time $t_{\text{f}}=560ps$; see \cite{EPAPS} for animated sequences. Interestingly, for $\beta = 7$ variant 3 is produced at intermediate times (Fig.\ \ref{fig:fig3}b) and vanishes at the end of the simulation. Indeed, the prescribed strain $\overline{\mathbf{F}}(t)$ is defined as an average of initial austenite [strain $\mathbf{I}$, volume fraction (VF) $1-t/t_f$], and of the two variants [strain $\mathbf{F}^\mr{M_1}$, VF $\alpha t/t_f$ and $\mathbf{F}^\mr{M_2}$, VF $(1-\alpha) t/t_f$]. However, the important energy gain when martensite forms (-13 meV/atom) favors larger fractions of both variants 1 and 2. Noting that the average strain produced by a combination of the three variants is null, this is balanced by a ``back'' strain proportional to $\mathbf{F}^\mr{M_3}$ inducing the formation of the third variant.

In the final state, the martensite compound forms a two-dimensional structure more complex than a simple laminate, in which the two possible orientations of twin interfaces consistent with boundary conditions at habitat planes \cite{NISH78} are simultaneously present. At meeting points of 90${}^\text{o}$-related interfaces, these boundary conditions cannot be satisfied and high elastic strains result. Relaxation occurs through a moderate formation of ``reversion'' austenite (less than 0.05\%) of the  $M_{1}\to A_2$ path, see Fig.\ \ref{fig:fig1}, represented as (barely noticeable) dark regions in Fig.\ \ref{fig:fig3}(a), right. Additional MD simulations and PF-RP calculations were made, using PBCs with planes rotated by an angle of 5$^{o}$. This small change inhibits chessboard patterning whatever the system size. A laminatelike structure takes place almost instantaneously without giving rise to any remarkable intermediate state. This suggests that stable chessboards may be difficult to observe experimentally for displacive transformations, even in small samples.

To conclude, we introduced a dynamic phase-field technique for martensitic transformations, fully compliant with crystal symmetries, that alleviates the need for vector Landau parameters, and obviously adaptative to a variety of situations. We illustrated it by an application to a model alloy. The good results obtained, which contrast with the crude approximations involved in modeling the energy landscape outside RPs, show that the details of these ``outer'' regions are most likely inessential to the main picture and confirm the relevance of a reaction-pathway approach to these questions. This view is supported by the agreement found with MD in size and time domains where both techniques could be compared, with a gain of two orders of magnitude in computational cost in favor of PF-RP.
\begin{acknowledgments}
C.D.\ and Y-P.P.\ thank G.\ Z\'erah and L.\ Truskinovski for seminal discussions.
\end{acknowledgments}


\end{document}